\documentclass[12pt]{article}
\newcommand{\BE}{\begin{equation}}
\newcommand{\EE}{\end{equation}}
\newcommand{\BA}{\begin{eqnarray}}
\newcommand{\EA}{\end{eqnarray}}

\newcommand{\doublespace}{\renewcommand{\baselinestretch}{1.45}\large\normalsize}
\doublespace
\input epsf.tex

\begin{document}

%\twocolumn[\hsize\textwidth\columnwidth\hsize\csname@twocolumnfalse\endcsname
\title{{\normalsize \bf Acoustic radiation in randomly-layered structures}}
\author{{\small Zhen Ye\thanks{Email: zhen@phy.ncu.edu.tw} \ and Pigang Luan}\\
{{\footnotesize \it Department of Physics, National Central
University, Chungli, Taiwan 32054, Republic of China}}}

\date{\small (\today)  }
\maketitle %\draft

\begin{abstract}

\noindent Radiation from acoustic sources located inside
randomly-layered structures is studied using the transfer matrix
method. It is shown that in contrast to the periodically layered
cases where the radiation can be either enhanced or inhibited
depending on the frequency, and the characteristics and the
material composition of the structures, in the random structures
the radiation is always inhibited. The degree of inhibition
depends on the acoustic frequency, number of random layers, and
the randomness and acoustic parameters of the structures. Both
spherically and cylindrically random structures are considered.
The results point to the possibility of designing sonic waveguide
devices which will not suffer from the energy loss caused by
radiation, thus allowing effective energy confinement or
long-range energy propagation.
\\
\vspace{1pt} \vspace{1pt}
\end{abstract}
%\pacs{PACS numbers: 43.20.Fn, 43.30.Gv., 43.30.Bp.}
%]

\section*{\small \bf INTRODUCTION}

When placed in spatially structured media, the radiation or
transmission of optical or acoustic sources will be modulated, a
fact of both fundamental importance and practical significance.
The structure-modulated transmission, called waveguide
propagation, is the backbone of modern opto-electronics and
acousto-optics systems. Designing proper waveguide devices that
can convey information without or with little energy loss thus has
been and continues to be a prime motivation for theoretical
studies of wave radiation and propagation in spatially structured
media \cite{Miller}.

Much effort has been focused on effects of metal and dielectric
interfaces, which can be constructed in either planar,
cylindrical, or spherical forms, on optical transmission and
radiation \cite{DK,TE,Ye1,Ye2,Science}. The optical transmission
or radiation in periodic structures has attracted particular
attention in different areas of applied physics, as in the
periodic situations the interaction between propagating waves and
structurally periodicity can be either constructive or
destructive, leading to significant enhancement or inhibition
respectively. In these situations, the waveguides act as a filter
which selects particular frequencies for propagation.
Understanding of optical propagation in periodicity has been vital
to the design of optical devices including optical fibers,
semiconductor lasers, hetero-junction bipolar transistors, quantum
well lasers, filters, and resonators
\cite{Miller,Science,Book1,Book2,Yab,Ping,M}.

Propagation of acoustic waves in spatially structured media has
also drawn attention recently \cite{Kush,Hawwa,Ye3,Ye4}. The
investigation of acoustic counterparts not only paves the way for
the possible design of new acoustic devices, but the acoustic
models themselves are advantageous in a number of situations.
First, as they are of scalar nature, the acoustic waves are
relatively simple to handle yet not to compromise the generality,
making them an ideal system for understanding more complicated
situations with vector wave propagation. For instance, the recent
results on acoustic propagation in water with parallelly placed
air-cylinders make it possible to study the ubiquitous phenomenon
of wave localization \cite{local} in an unprecedented detailed and
manageable manner \cite{Ye4,Emile}; the localization of optical
waves has posed a long standing problem and a subject of much
debate \cite{Stat}. Second, the research is motivated by potential
applications in acousto-optic fiber devices\cite{APL}. Third, it
is relatively easy to manufacture hetero-structures with large
contrast in acoustic impedance. This allows not only the study of
strong scattering, but the usage of the properties of the strong
scattering in such situations as noise reduction\cite{tba2}.

In the previous paper \cite{Ye4}, the acoustic radiation from a
source located inside periodically layered spherical cavities has
been considered. It is found that significant enhancement or
inhibition on the radiation is possible by varying the acoustic
parameters and the periodicity of the structures of the guides.
The analysis predicts well-defined peaks and nodes in the
cavities. The fact that wave transmission in the periodical
situations is only possible for certain ranges of frequency is
useful as it is of help for devising apparatuses such as filters.
For such applications such as energy transport, however, it is
desired that no or little energy be radiated for any frequencies.
In other words, devices are designed so that energy are confined
inside the devices. In this article, we consider acoustic
radiation from acoustic sources located inside randomly layered
structures. The structures can be of spherical and cylindrical
geometries. We show that the radiation is inhibited for all
frequencies for any given randomness, and waves are confined in
the structures. The confinement or localization effect becomes
increasingly significant as the randomness increases or the
acoustic contrast increases for a given randomness. These features
hold for both spherical or cylindrical structures. The results may
be of help in the design of wave processing devices such as
resonators or efficient energy guides.

\section{\small \bf FORMULATION OF THE PROBLEM}

We consider a unit acoustic source located insider a layered
structure, which can be either spherical or cylindrical geometry.
The conceptual layout is sketched in Fig.~1. The most inner and
outer radii of the structure is $R_1$ and $R_N$ respectively.
Between $R_1$ and $R_N$, there are $N-1$ randomly placed
interfaces. The boundary at $R_1$ ($i=1,2,\dots, N$) is denoted as
the $i$th interface. These $N-1$ interfaces are {\it randomly}
located between $R_1$ and $R_N$. The further parameters are as
follows. The sound speed and the mass density inside $R_1$ are
$c_1$ and $\rho1$, while the sound speed and mass density of the
surround medium, i.~e. outside $R_N$, are denoted by $c$ and and
$\rho$ respectively. The sound speed and mass density between the
$i$th and the $(i+1)$th interfaces are $c_{i+1}, \rho_{i+1}$. We
define $g_i = \rho_i/\rho$ and $h_i=c_i/c$. The parameters $g_i$
and $c_i$ are called acoustic contrast parameters.

The Helmholtz wave equation inside the structure is \BE
(\nabla^2+k_1^2)p(\vec{r}) = -4\pi\delta^{(d)}(\vec{r}), \ \
\mbox{for} \ r<R_1, \label{eq:1}\EE where $\nabla^2$ is the usual
Laplacian operator, $k_1$ is the wave number ($k_1 =\omega/c_1$),
and $\delta$ is the Dirac delta function representing the source.
The superscript $d$ refers to the dimension. For spherical
structures, $d=3$, while for cylindrical structures $d=2$. The
wave equation outside the structure is \BE
(\nabla^2+k^2)p(\vec{r})=0, \ \ \mbox{for} \ r>R_N,
\label{eq:2}\EE with $k=\omega/c$. Similarly, the wave equations
for the random media between $R_1$ and $R_N$ are \BE
(\nabla^2+k_i^2)p(\vec{r}) = 0, \ \mbox{for} \ i=1,2,3,\dots, N-1,
\ \mbox{and} \ R_i<r<R_{i+1}, \label{eq:3} \EE with $k_i =
\omega/c_i$.

Before going into details of the problem we make a detour to
consider a more general transmission through an arbitrary
interface. The general solution to wave equation such as those in
Eqs.~(\ref{eq:1}), (\ref{eq:2}), and (\ref{eq:3}) can be written
as \BE p_i(r) = A_i G^{(d)}(k_ir) + B_i{G^{(d)}}^\dagger(k_ir), \
(i=1,2,\dots), \EE where ``$\dagger$" means taking the complex
conjugate, $A_i$ and $B_i$ are coefficients to be determined by
boundary conditions, and $G^{(d)}$ is the Green's function in the
$d$-th dimension and is written as \BE G^{(d)}(k_ir) =
\left\{\begin{array}{ll} i\pi H_0^{(1)}(k_ir), & d = 2\\
k_i\frac{e^{ik_i r}}{k_ir}, & d =3.\end{array} \right.\EE Note
that the Green's function $G^{(d)}$ also represents the wave
transmitted from the unit source without the presence of the
layered structures; its complex conjugate represents the inward
moving wave.

To solve for the unknown coefficients $A_i$ and $B_i$, we invoke
the boundary conditions that require the pressure field and the
radial displacement be continuous across the interfaces. Consider
an arbitrary interface at $R$. The wave on the inner ($<$) and
outer ($>$) sides of the interface are respectively denoted by \BE
p_{<,>}(r) = A_{<,>}G^{(d)}(k_{<,>}r) +
B_{<,>}{G^{(d)}}^\dagger(k_{<,>}r).\EE The coefficients $A$ and
$B$ on the two sides of the interface can be related by the usual
boundary conditions, \BE A_{<}G^{(d)}(k_{<}R) +
B_{<}{G^{(d)}}^\dagger(k_{<}R) = A_{>}G^{(d)}(k_{>}R) +
B_{>}{G^{(d)}}^\dagger(k_{>}R), \label{eq:bc1} \EE and \BE
\frac{k_<}{\rho_<}\left[A_{<}{G^{(d)}}^\prime(k_{<}R) +
B_{<}{G^{(d)}}^{\dagger,\prime}(k_{<}R)\right] =
\frac{k_>}{\rho_>}\left[A_{>}{G^{(d)}(k_{>}R)}^\prime +
B_{>}{G^{(d)}}^{\dagger,\prime}(k_{>}R)\right],\label{eq:bc2} \EE
where ``$\prime$" means the derivative, e.~g. ${G^{(d)}}^\prime(x)
= \frac{d}{dx}G^{(d)}(x).$ Eqs.~(\ref{eq:bc1}) and (\ref{eq:bc2})
can be written in the matrix form \BE \left(\begin{array}{c} A_<
\\ B_<\end{array}\right) = T(R) \left(\begin{array}{c} A_>
\\ B_>\end{array}\right),\EE with
\BE T(R) = \left(\begin{array}{cc} G^{(d)}(k_<R) &
{G^{(d)}}^\dagger(k_<R)\\ \frac{k_<}{\rho_<}{G^{(d)}}^\prime(k_<R)
& \frac{k_<}{\rho_<}
{G^{(d)}}^{\dagger,\prime}(k_<R)\end{array}\right)^{-1}
\left(\begin{array}{cc} G^{(d)}(k_>R) & {G^{(d)}}^\dagger(k_>R)\\
\frac{k_>}{\rho_>}{G^{(d)}}^\prime(k_>R) &
\frac{k_>}{\rho_>}{G^{(d)}}^{\dagger,\prime}(k_>R)\end{array}\right).
\EE The matrix $T$ is called the transfer matrix.

For a system consisting of multiple interfaces, the transmission
and reflection coefficients can be related through a consecutive
product of the transfer matrices at all the interfaces. We denote
the resulting matrix by ${\bf M}$. Therefore, the waves inside and
outside the layered structure are related through \BE
\left(\begin{array}{c} A_{in}
\\ B_{in}\end{array}\right)= {\bf M}
\left(\begin{array}{c} A_{out}
\\ B_{out}\end{array}\right),
\label{eq:m}\EE where $${\bf M} = \prod_{i=1}^N T(R_i) =
\left(\begin{array}{cc}m_{11} & m_{12}\\ m_{21} &
m_{22}\end{array}\right). $$

Now we come back to the problem. Inside the most inner interface,
the transmitted wave from the source is subject to reflection from
the interface at $R_1$. The total wave can be expressed as \BE
p(r) = G^{(d)}(k_1r) + p_R(r), \ \ \mbox{for} \ 0<r<R_1,\EE where
the first term is the transmitted wave from the source, and the
second term is the reflected wave. Since the reflected wave must
be finite at the origin, it can be written as \BE p_R(r) =
-QG^{(d)}(k_1r) + Q {G^{(d)}}^\dagger(k_1r). \EE The negative sign
in front of the first term asserts that the reflected wave remains
finite at the origin. Thus the total wave inside the layered
structure is \BE p(r) = (1-Q)G^{(d)}(k_1r) + Q
{G^{(d)}}^\dagger(k_1r),\label{eq:14}\EE which gives $A_{in} =
1-Q$ and $B_{in} = Q$. Another observation is that there is no
reflected wave outside the layered structure, i.~e. beyond $R_N$,
we have \BE B_{out}=0, \ \ \mbox{for} \ r>R_N.\EE Taking these
into consideration, Eq.~(\ref{eq:m}) becomes \BE
\left(\begin{array}{c} 1-Q \\ Q\end{array}\right) =
\left(\begin{array}{cc}m_{11} & m_{12}\\ m_{21} &
m_{22}\end{array}\right) \left(\begin{array}{c} A_{out}
\\ 0\end{array}\right). \EE
This equation yields the solutions \BE
Q=\frac{m_{21}}{m_{11}+m_{21}}, \ \ \
A_{out}=\frac{1}{m_{11}+m_{21}}. \EE

>From these solutions, the radiated acoustic intensity can be
computed from \BE I_{out}(r) = \frac{1}{2\rho c}
\left(\frac{2\pi}{k}\right)^{3-d}\frac{|A_{out}|^2}{r^{d-1}}.\EE
The reflected intensity is \BE I_{R}(r) = \frac{1}{2\rho_1
c_1}\left(\frac{2\pi}{k_1}\right)^{3-d}\frac{|Q|^2}{r^{d-1}}.\EE
>From Eq.~(\ref{eq:14}) states that the effective source
transmission intensity is \BE I_s(r) = \frac{1}{2\rho_1
c_1}\left(\frac{2\pi}{k_1}\right)^{3-d} \frac{|1-Q|^2}{r^{d-1}}.
\EE Energy conservation states \BE
\left(\frac{2\pi}{k_1}\right)^{3-d}\frac{1}{\rho_1 c_1}|1-Q|^2 =
\left(\frac{2\pi}{k}\right)^{3-d}\frac{1}{\rho c}|A_{out}|^2 +
\left(\frac{2\pi}{k_1}\right)^{3-d}\frac{1}{\rho_1c_1}|Q|^2.
\label{eq:ec}\EE We note some typographical errors about
Eqs.~(21), (22) and (23) in \cite{Ye3}.

We define the transmission (TR) and reflected (RF) coefficients as
follows \BE \mbox{TR} = |A_{out}|, \ \ \mbox{RF} = |Q|. \EE It
will become clear that when randomness is added, the transmission
will be inhibited for all frequencies. The energy will be confined
inside the layered structure. With the random layers, the energy
flow in the radial direction decreases as the randomness or the
number of layers increases, a useful property for random layered
cavities to transport energies.

\section{\small \bf NUMERICAL RESULTS}

The situation that the coating layers are periodically placed for
spherical cavities has been considered in \cite{Ye3}. For the
reader's convenience, we re-plot the results in Fig.~\ref{fig2}.
In the computation, the layered structure is constructed as ${\bf
A-M-W-M-W-\cdots-W}$, where ${\bf A}$ represents the air inside
the cavity, i.~e. the air fills the space $r<R_1$, ${\bf M}$
refers to the coating material whose acoustic impedance is $\rho_m
c_m$, and $W$ refers to another coating material and we assume it
to be similar to that of the surrounding medium, i.~e. the water.
We define $g=g_2 = \rho_m/\rho$ and $h=h_2=c_m/c$. Here we take
the acoustic contrasts as $\rho_m/\rho = 2, c_m/c=2$. The
parameters for air are $\rho_1/\rho = 0.00129, \ c_1/c = 0.23$.
The horizontal bars represent the interfaces. There are 30
interfaces. The thickness of each layer is set to be identical and
equals $0.2R_1$; thus the total thickness of the layered materials
is $29\times 0.2R_1$.

Figure~\ref{fig2} indicates the following. (1) Periodically
layered structures selects particular frequencies for
transmission, i.~e. at these frequencies, the transmission is
greatly enhanced. (2) The spectral valleys in which transmission
is greatly inhibited are equivalent to the forbidden bands
observed in regular lattice solids. These valleys are called
forbidden frequency bands. It is known from the previous results
that these forbidden bands are caused by Bragg reflection for a
periodic structure\cite{Ye3}. (3) The reflection is also
significantly enhanced at certain frequencies, and the separation
of the reflection peaks is almost constant. (4) It is interesting
to see that the results for both spherical and cylindrical
geometries are qualitatively similar. The transmission and
reflection peaks are only shifted slightly. In the later
discussion, we will focus on the results for the cylindrical
geometry while the relevant results for the spherical structure
will be added only when needed.

Now we consider the randomly layered cases. In the simulation, the
layered structures are constructed in the similar way as in the
periodical structures, except that the interfaces are randomly
placed. To be explicit, the layered structure is ${\bf
A-M-W-M-W-\cdots-W}$. There are $N$ interfaces, i.~e. there are
$N$ horizontal bars. When there is no randomness, the interfaces
are equally spaced along the radial direction; the distance
between the interfaces is $D=0.2R_1$. In this way, the $i$-th
interface is located at $r_i=R_1+(i-1)D$ with $i$ ranging from 1
to $N$. The level of randomness is controlled by the degree of
allowing the interfaces to shift from their locations for
underlying periodically structures. We define the randomness as
$\Delta=\frac{|\delta|}{D}$, where $\delta$ is the range within
which the interfaces are allowed to shift from their locations
when there are no disorders. For example, the location of the
$i$-th interface can be randomly varied with the range between
$R_1+(i-1)D - \delta$ and $R_1+(i-1)D + \delta $. Clearly, the
total random is the case that $\Delta =1$.

The effects of the level of randomness, numbers of random layers,
and acoustic contrast on the radial transmission are shown in
Fig.~\ref{fig3}. As both spherical and cylindrical geometries have
the similar features, we only plot the results for the cylindrical
structures here. The important message from the figure is that
when the randomness is added, the transmission is repressed for
all frequencies. In other words, no energy can escaped and all
energies are confined inside the structure. In particular,
Fig.~\ref{fig3} shows the following. (1) When the randomness is
introduced, the transmission becomes subdued; when the randomness
is small, the band effects from the underlying periodic structures
are still noticeable for low frequency bands. This is shown by the
case of $\Delta = 0.3$. (2) The inhibition is more significant for
high frequencies. For the low frequencies, the inhibition seems
still possible. This is due to the finiteness in the number of
layers. With increasing number of layers, the inhibition will be
extended to low frequencies. (3) Fixing the number of layers and
when the randomness is greater than a certain value, the effect of
the variation in randomness becomes not prominent for high
frequencies. This is shown by the tendency that the curves for
$\Delta = 0.3$ and $1$ merge at high frequencies, referring to the
case in, for example, (c) and (e). (4) Increasing the number of
random layers, the transmission will be reduced further. In fact,
the transmission will decay exponentially with the number of the
random layers, as will be shown later. (5) The above features hold
when the acoustic parameters vary. (6) For the spherical geometry,
we obtain the similar results as in Fig.~\ref{fig3}, and because
of this we do not show the results here.

With increasing the number of the random layers, the transmission
will decrease exponentially. This is illustrated by
Fig.~\ref{fig4} for the case $g=h=2$. Here the results are
averaged over 200 random configurations. Here is shown the
transmission versus the number of random layers for four
frequencies and four randomness levels. Out of the four
frequencies, two are located at where the transmission is possible
when no randomness is introduced, and one is within and one is
within but close to the edge of the forbidden band of the
underlying periodic structure. It is clear that the increasing
randomness decreases gradually the transmission in the regimes in
which the transmission is possible when there is no disorder.
These regimes may be called the passing bands for the underlying
periodical structures. This property is also true for frequencies
with but in the vicinity of the forbidden band. Inside the
forbidden bands, however, though inhibited the transmission is
increased with the added disorder as compared to the case without
disorders. This property has also been observed in one-dimensional
random liquid media\cite{Luan,Maradudin}. This indicates that the
inhibition due to randomness and the inhibition due to Bragg
reflection may have different causes, and the two effects compete.
For all frequencies, be within the passing or forbidden bands, the
transmission decreases exponentially with increasing numbers of
random layers, implying that the most energy is localized near the
transmitting source.

The relation between the transmission and the number of layers can
be approximately described as \BE I= \frac{I_0}{r^{d-1}}
e^{-2N/N_0}. \EE The parameter $N_0$ represents the effective
number of random interfaces to localize the energy, and is named
the localization length in terms of the number of interfaces. The
localization lengths for the cases in Fig.~\ref{fig4} are
summarized in Table~\ref{table1}. It is clear from the table that
within the passing bands of the underlying period structure the
localization length decreases with increasing frequency and
disorders, while well within the forbidden bands the localization
length decreases with increasing disorders. This table also shows
that the localization lengths are almost identical for the
spherical and cylindrical geometries. Moreover, we also see that
for the totally random configurations, the energy confinements can
be achieved by just a few random interfaces. Further simulation
indicates that the localization length decreases with increasing
the acoustic contrast.

\begin{table}[hbt]
\caption{\label{table1} Localization length vs $kR_1$ for various
randomness $\Delta.$} %\vspace{10pt}
\begin{center}
\begin{tabular}{cc} \hline \hline
\hspace{24mm}Spherical Shell\hspace{20mm}Cylindrical Shell
$\mbox{\hspace{7mm}}$\\ \hline \hline
\end{tabular}
\begin{tabular}{cccccccccc}
$kR_1\backslash\Delta$ & 0.1 & 0.3 & 0.5 & 1.0 & & 0.1 & 0.3 & 0.5
& 1.0 \\ \hline
 5.5 & 355.7 &  42.9 & 18.0 & 7.7 & & 390.3 & 44.2 & 17.9 & 7.7\\
 6.0 & 145.2 & 20.8 & 11.0 & 6.2 & & 143.7 & 20.4 & 11.1 & 6.3\\
 6.5 & 12.4 & 8.8 & 6.9 & 5.5 & & 12.4 & 8.8 & 7.0 & 5.4\\
 8.0 & 2.0 & 2.1 & 2.7 & 4.5 & & 2.0 & 2.1 & 2.7 & 4.5\\
\hline
\end{tabular}
\end{center}
\end{table}

We also examine the reflection behavior. The conservation law in
Eq.~(\ref{eq:ec}) is rewritten as
\BE\left(\frac{2\pi}{k_1}\right)^{3-d} \frac{1}{\rho_1
c_1}\left(|1-Q|^2-|Q|^2\right) =
\left(\frac{2\pi}{k}\right)^{3-d}\frac{1}{\rho c}|A_{out}|^2. \EE
When the radiation is stopped, i.~e. $|A_{out}|\rightarrow 0$, we
are thus led to the relation \BE |1-Q|^2\approx |Q|^2. \EE The
numerical confirmation is shown in Fig.~\ref{fig5}, in which the
ratio $|1-Q|/|Q|$ is plotted as a function of frequency in terms
of $kR_1$. With reference to Fig.~\ref{fig2}, we see that the
ratio deviates from one only for frequencies within the passing
bands for the regular structures. With added disorders, this ratio
is virtually one for all frequencies. Note that the reason why the
ratio is close to one even for the periodic structures is that we
take out a term associated with the acoustic impedance.

Finally we note that the results in this papers bear some
similarities to the Anderson localization in one dimensional
random systems in that no waves can propagate in such a
system\cite{Luan}. The most significant difference is that in the
present case, the waves are localized near the transmitting site.
In the 1D cases, however, the energy needs not be confined near
the source, and there is a stochastic resonance behavior which is
absent from the present situations.

\section{\small \bf SUMMARY}

In this paper, we consider acoustic radiation from randomly
layered cavities. The results in the present paper convey the
information that when a cavity is coated by random layers,
virtually no energy can be radiated in the radial direction. The
waves are mainly localized inside the cavity, and its decay along
the radial direction follows an exponential law. The results
presented here may be useful for designing acoustic `lasers',
resonators, or energy transport. For example, suppose there is a
cylindrical wave guide. The propagating wave can be generically
written as $e^{ik_zz}f(r)$ in the cylindrical coordinates. When
the guide is coated by random materials, according to the results,
no energy can be radiated into the transverse directions, all
energies will be confined inside the guide and can only propagate
in the longitudinal direction, i.~e. $f(r)$ decreases with $r$. By
adjusting the coating materials, the energy transport along the
longitudinal axis may be tuned to fit the applications. The
results may also be useful for designing possible `sonic fibers'
in analogy with the recent all-dielectric optical fibers
\cite{Science}. For the spherical waveguides, various energies can
be stored inside the guides by adjusting the coating materials.

\section*{\small \bf ACKNOWLEDGEMENT}

The work received support from the National Science Council.

%\begin{references}

\newpage

\begin{figure}
\begin{center}

\epsfxsize=6in\epsffile{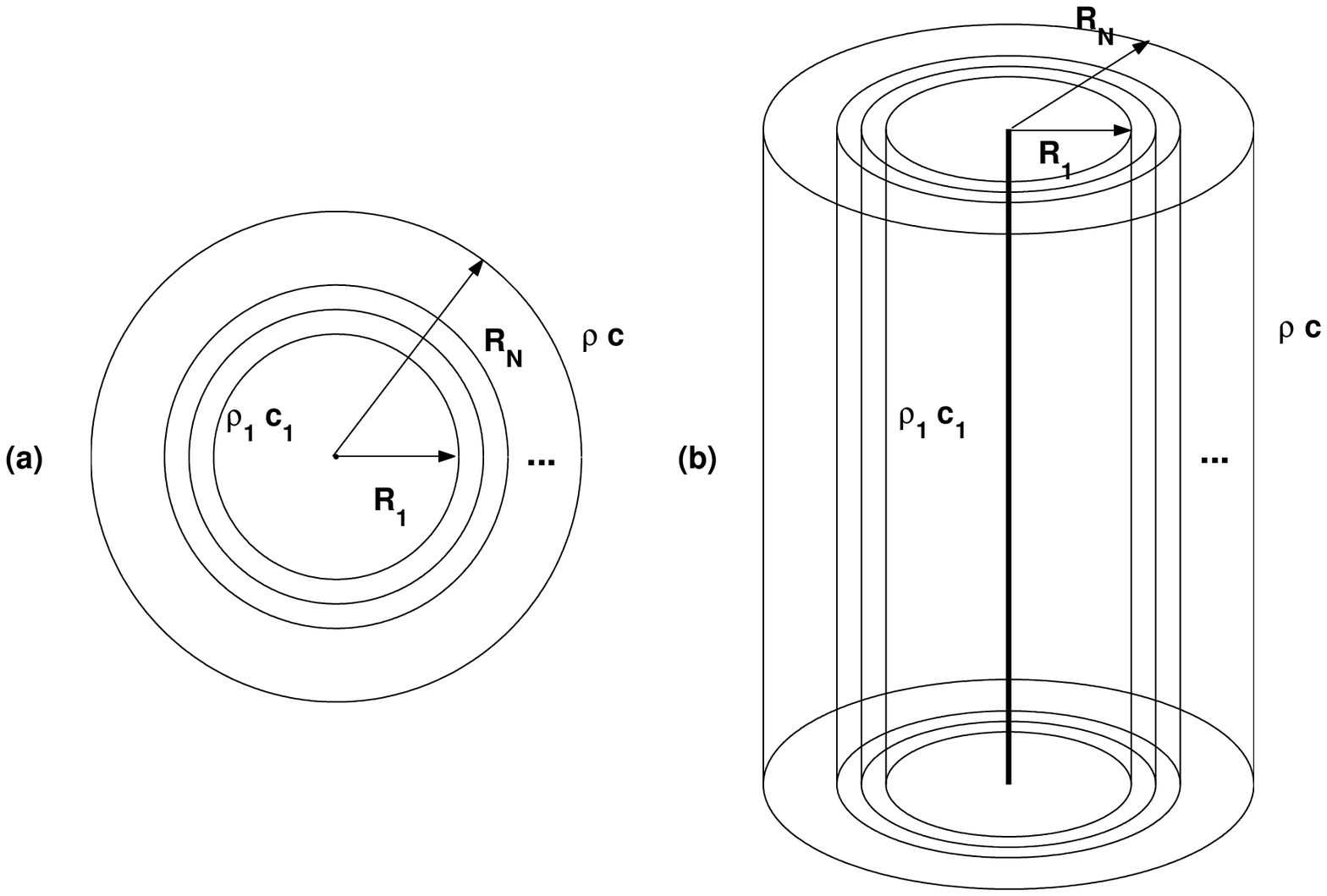}
\vspace{14pt}\caption{Conceptual layout for spherical and
cylindrical cavities.} \label{fig1}
\end{center}

\end{figure}

\begin{figure}
\begin{center}
\epsfxsize=5in\epsffile{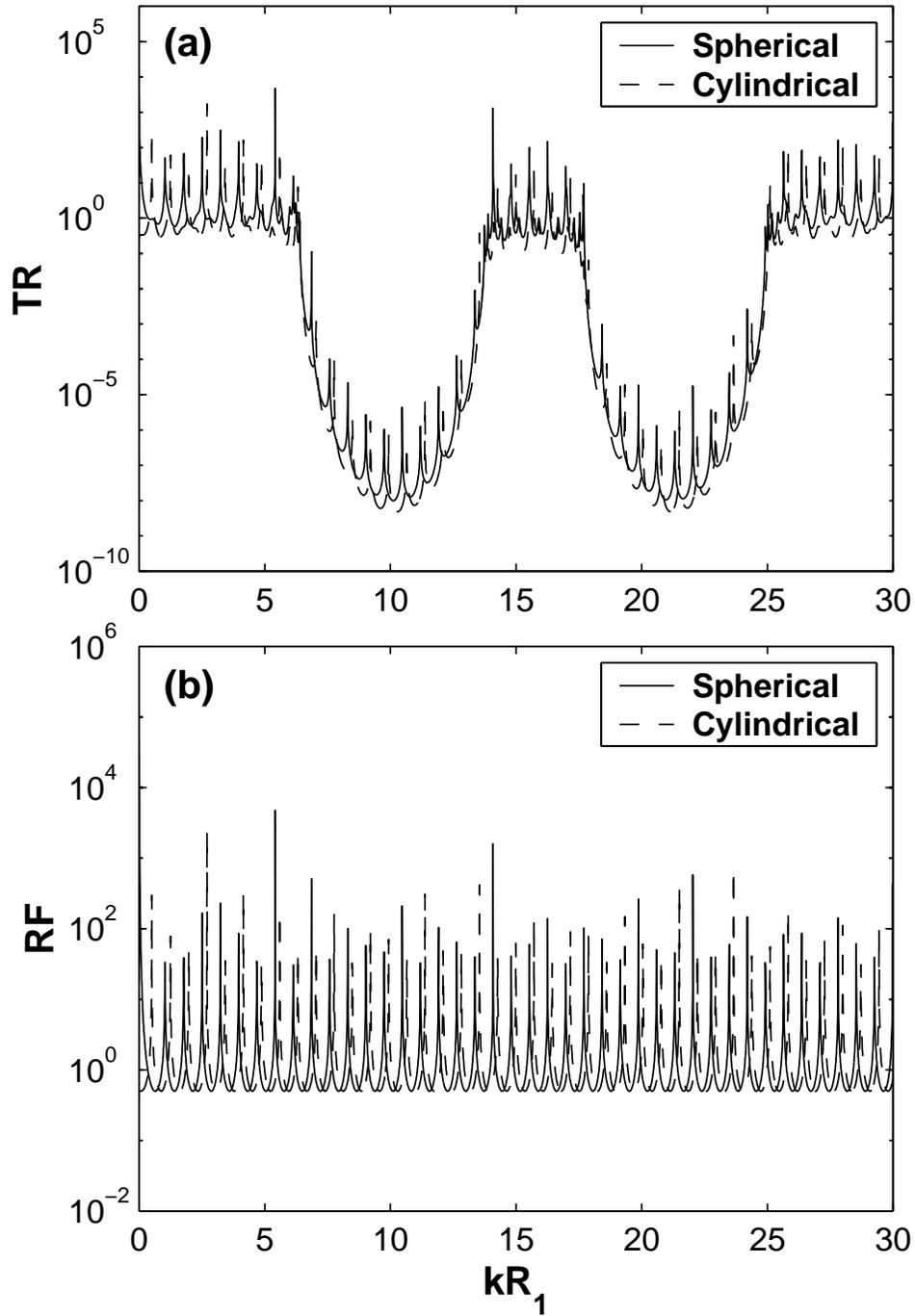}
\vspace{14pt}\caption{Transmission and reflection coefficients as
a function of frequency in terms of $kR_1$ for periodically
layered cavities: (a) Transmission; (b) Reflection. } \label{fig2}
\end{center}
\end{figure}

\begin{figure}
\begin{center}
\epsfxsize=5in\epsffile{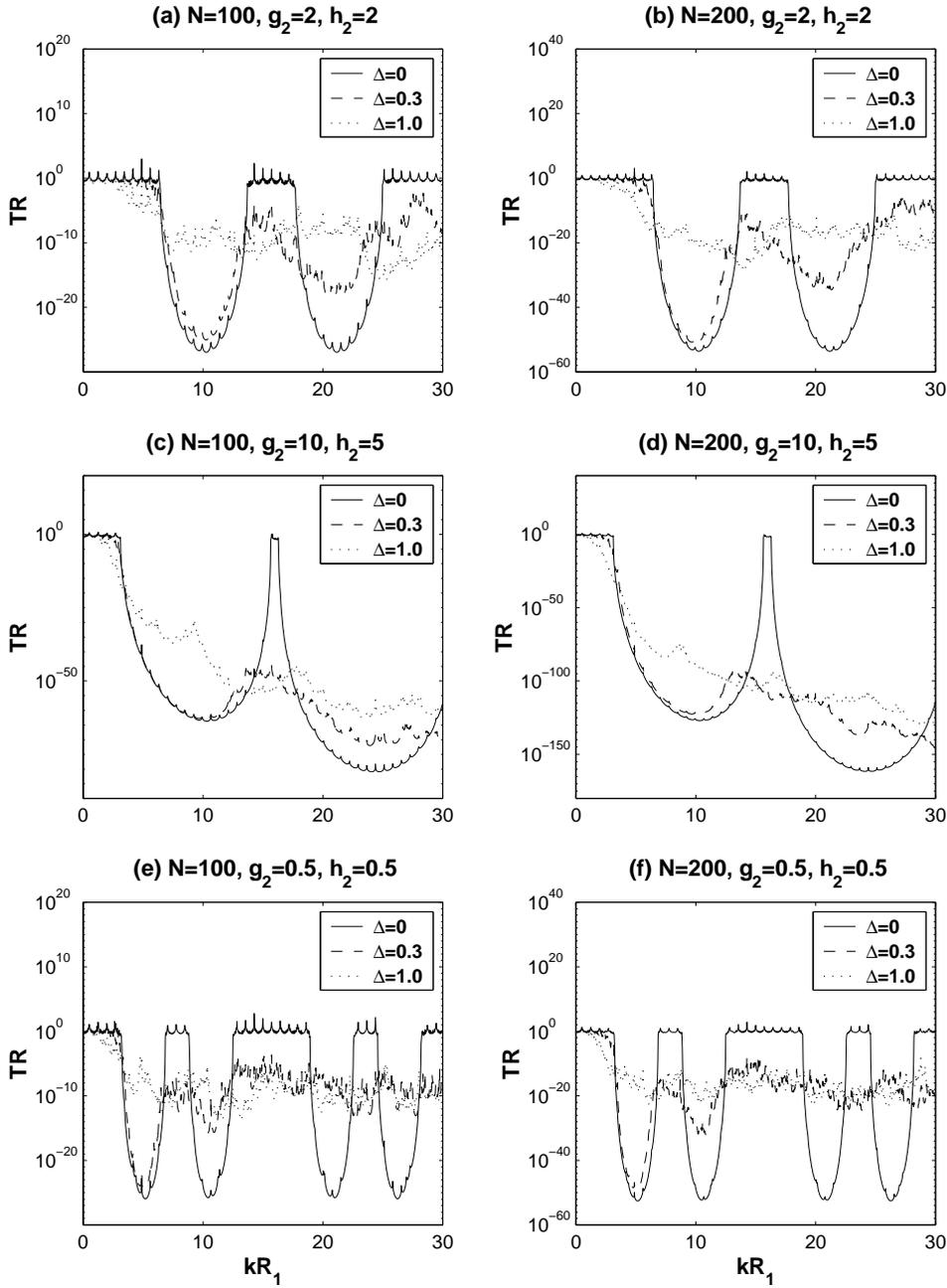}
\vspace{14pt}\caption{Transmission as a function of frequency in
terms of $kR_1$ for randomly layered cylindrical structures for
various disorders, acoustic contrasts, and numbers of random
layers.The results shown are for an arbitrary random
configuration.} \label{fig3}
\end{center}
\end{figure}

\begin{figure}
\begin{center}
\epsfxsize=5in\epsffile{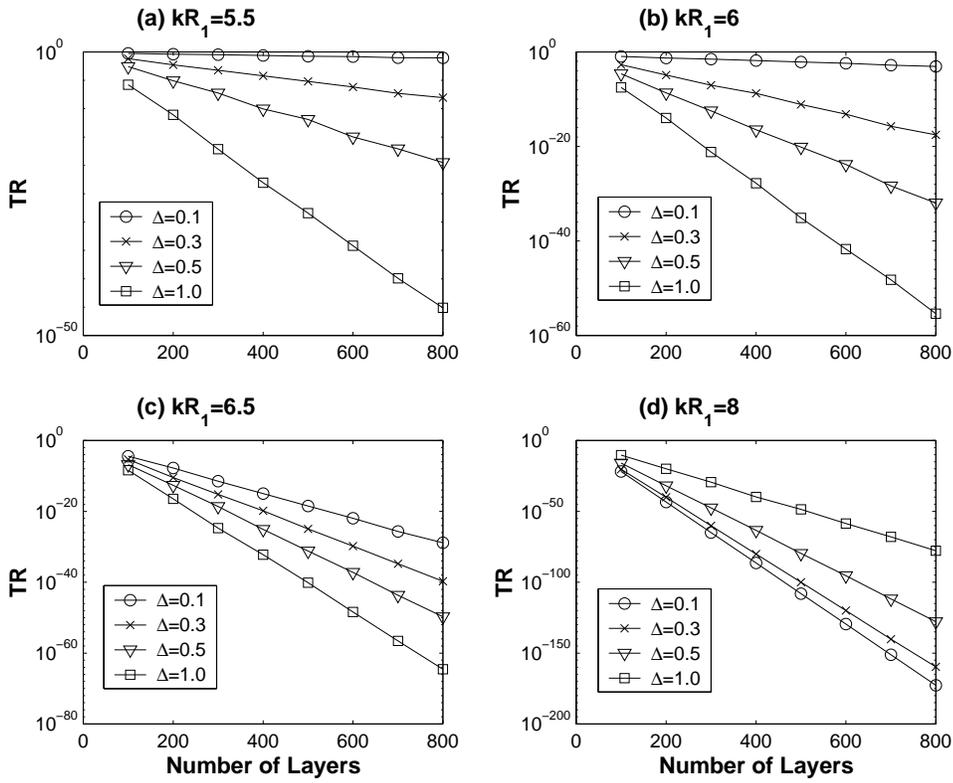}
\vspace{14pt}\caption{Transmission as a function of the number of
random layers for various frequencies and different randomness}
\label{fig4}
\end{center}
\end{figure}

\begin{figure}
\begin{center}
\epsfxsize=5in\epsffile{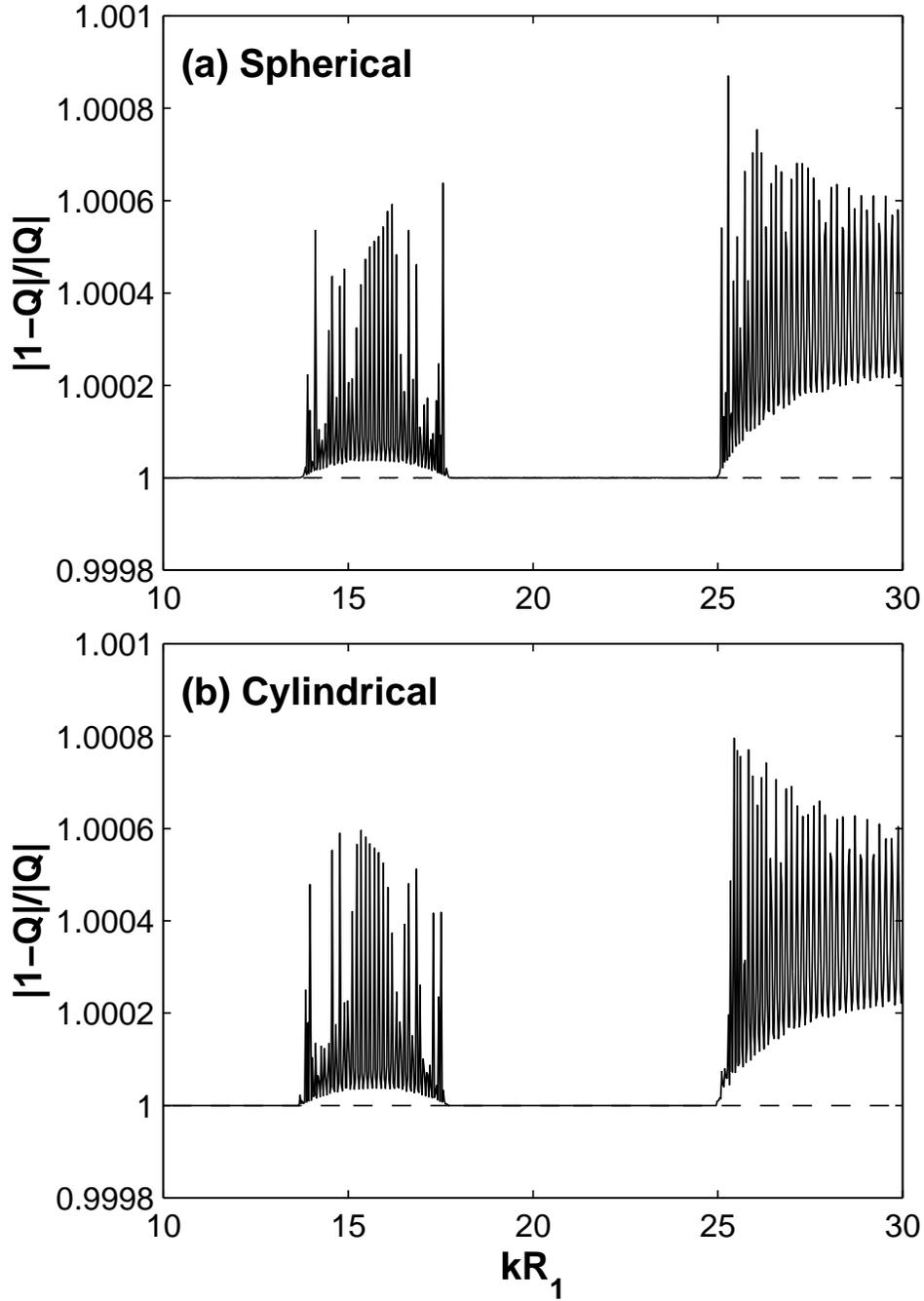} \vspace{14pt}\caption{The ratio
$|1-Q|/|Q|$ versus $kR_1$. The solid and dashed lines are for the
periodic and random structures respectively. Here $g=h=2$ and
$N=100$} \label{fig5}
\end{center}
\end{figure}

\end{document}